\documentclass{dasc}

\usepackage[english]{babel}
\usepackage{fixltx2e}
\usepackage[rflt]{floatflt}
\usepackage{adjustbox}
\usepackage{graphicx}
\usepackage{color}
\usepackage{url}
\usepackage{spverbatim}

\usepackage{trackchanges}
\title{\uppercase{Challenges of Security and Trust of Mobile Devices as Digital Avionics Component}}

\begin{document}

\author{
	Raja Naeem Akram, Konstantinos Markantonakis, Royal Holloway, University of London. Egham, UK \\
}

\maketitle

\section*{Abstract}

Mobile devices are becoming part of modern digital avionics. Mobile devices can be applied to a range of scenarios, from Electronic Flight Bags to maintenance platforms, in order to manage and configure flight information, configure avionics networks or perform maintenance tasks (including offloading flight logs). It can be argued that recent developments show an increased use of personal mobile devices playing an integral part in the digital avionics industry. In this paper, we look into different proposals for integrating mobile devices with various avionics networks -- either as part of the Bring Your Own Device (BYOD) or Corporate Owned Personally Enabled (COPE) paradigms. Furthermore, we will evaluate the security and trust challenges presented by these devices in their respective domains. This analysis will also include the issues related to communication between the mobile device and the aircraft network via either wired or wireless channels. Finally, the paper puts forward a set of guidelines with regards to the security and trust issues that might be crucial when enabling mobile devices to be part of aircraft networks.

\section*{Introduction}
In the aviation industry, there is a growing proliferation of mobile devices, including tablets, smart phones and portable computers. These devices are increasingly used not only by the passengers but also by aircraft crew like pilots and maintenance staff. For example, a pilot can create a flight plan on his or her personal tablet (Electronic Flight Bag) and then upload it to the aircraft \cite{Freem2012}. In this scenario, either the aircraft manufacturer or the airline provides the pilot with the necessary software application to perform this task \cite{Carrico2015}. 

In addition to mobile devices being used by the on-board aircraft crew, they are also used by off aircraft maintenance crew for downloading or uploading necessary data from the aircraft. 

Any mobile devices, whether used by on-board or off-board crew, have to interface with the aircraft, using either a wired or a wireless interface. These interfaces present their own unique sets of security and operational issues. 

It is projected that such devices might bring operational benefits to pilots, on-board crew members and maintenance crews. At the same time, it is documented that use of such technologies might increase automation bias, complacency, aircrew distraction and potential software errors \cite{Johnstone2013}. In addition to the listed concerns, these devices potentially create additional security and reliability issues. Depending upon the device, there is a potential that a compromised mobile device might give a malicious entity a route to an aircraft's on-board computers --- and depending upon the function of these devices the malicious entity could cause damage.

Another complication is who owns the mobile device, and this ownership might restrict what an aviation organisation (airline, or aircraft manufacturer) can enforce. It is challenging to force a user to follow a particular policy on a device that basically belongs to them. This is a traditional security problem regarding users' limitations in following a policy, and therefore users are usually considered the weakest link in a security-related process or framework. 

In the context of this paper, we define a mobile device as an off-the-shelf portable computing device that can be used by pilots, aircraft crew, or maintenance personnel. These devices can display, store, process and communicate related information -- in on-aircraft systems and off-aircraft systems alike, using either a wired or wireless medium. Examples of such devices include smart phones, tablets and potentially laptop computers.

\begin{figure}[htbp]
	\centering
		\includegraphics[width=1.00\columnwidth]{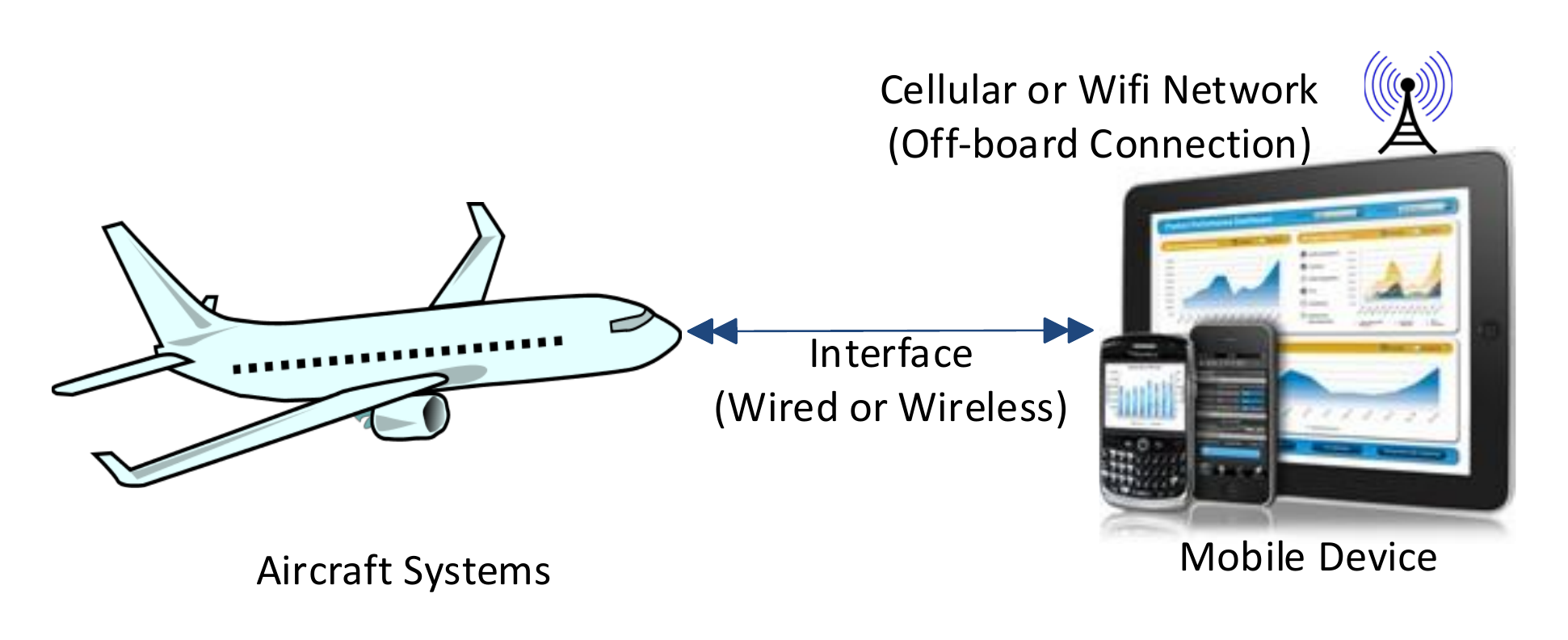}
	\caption{Mobile Device Connectivity with Aircraft Systems}
	\label{fig:MobileDevicesAircraft}
\end{figure}

Figure \ref{fig:MobileDevicesAircraft} illustrates mobile devices potentially connected to aircraft systems either by a wired or wireless connection, while some of the mobile devices might also have either cellular or Wi-Fi capability to connect to off-aircraft systems (i.e. airline networks, aircraft manufacturer networks etc.). It must be noted that in this paper we will focus solely on mobile devices and not ``installed devices'' such as the Class 3 Electronic Flight Bags (EFB) defined in Federal Aviation Administration (FAA) Advisory 120-76 \cite{FAA120-76C}. 

In this paper, we discuss different operational frameworks for incorporating mobile devices into on-board digital avionics, how these mobile devices can interface with the aircraft system(s), what security and reliability issues each of these schemes might present and finally how to resolve security issues, along with examining the technologies that might be used to safeguard mobile devices.

\begin{table*}[t]
	\centering
	\begin{adjustbox}{max width=\textwidth}
		\begin{tabular}{|l|c|c|c|}
		\hline
	{\bf Criteria} & {\bf COPE} & {\bf BYOD} & {\bf CYOD} \\ \hline
	Device Ownership & Company (IT Department) & Employee (User) & Company's Approved/Preconfigured Devices\\ \hline
	Application Control & Full (IT Department) & Full (User) &  Partial (IT Department and User) \\ \hline
	Protecting Company Assets & Full (IT Department) & Partial (User) & Partial (IT Department and User) \\ \hline
	Responsibility for Securing the Device & IT Department & User & Partial (IT Department and User)\\ \hline
	Users' Privacy Issues & Significant & Limited  & Limited \\ \hline
	Usability and Freedom of Use for Employees & Limited & Full & Limited \\ \hline	
		\end{tabular}
	\end{adjustbox}
	\caption{Comparison between COPE, BOYD and CYOD}
	\label{tab:ComparisonBetweenCOPEBOYDAndCYOD}
\end{table*}

\section*{Corporate Integration Mobile Device Paradigms}

Two major paradigms to either incorporate mobile devices in an organisational computer system or to integrate mobile devices with an aircraft's on-board avionics systems are COPE and BYOD. In this section, we will discuss COPE and BYOD, along with wired and wireless interfaces these devices can use to communicate with avionics systems. Table \ref{tab:ComparisonBetweenCOPEBOYDAndCYOD} provides a comparison between the different integration paradigms that a corporate can deploy to bring in mobile devices. A detailed overview is provided in the following sections that explain the elements in the Table \ref{tab:ComparisonBetweenCOPEBOYDAndCYOD}.

\subsection*{Corporate Owned Personally Enabled (COPE)}

In this paradigm, a corporate entity, for example an airline, aircraft manufacturer or any other avionics-related company, acquires mobile devices and customises these devices to suit its security needs \cite{walters2013bringing}. After installing the management software and any other applications necessary for employees to perform their jobs, these devices are then given to individual employees. By acquiring the devices for employees, the corporation can make sure that only high quality and secure devices connect to their aircraft system. In addition to this, the devices allow users (employees) to also use them as they desire -- by enabling them to download applications they prefer. Furthermore, as the devices are owned by the corporate entity, they can easily manage their enrolment  as secure devices to connect and communicate with the aircraft systems.

Furthermore, the IT department of the corporation can keep a track of its mobile devices. If required, the IT department can push updates and also prohibit applications that are deemed dangerous to the secure functioning of their applications. In this paradigm, the corporate entity can exercise strong control over its mobile devices. However, at the same time, employees might not get the same level of usability they enjoy on their personal devices. Furthermore, if the mobile device was acquired off-the-shelf then the underlying platform and hardware would still be out of the company's control. Any loopholes in the underlying platform and hardware would require repair and updating by the mobile device manufacturer. A brief explanation \ref{tab:ComparisonBetweenCOPEBOYDAndCYOD} in the context of COPE is given below:

\begin{itemize}
	\item Device Ownership: As discussed before, the company acquires the devices and configures them as it desires before issuing them to their employees. The company has full control and in turn full responsibility for security in this model. 
	\item Application Control: As the device is under the full control of the company, the company can install or delete any application it desires. Furthermore, it can also prohibit certain applications from being installed on to their devices by their respective users (i.e. employees). 
	\item Protecting Company Assets: As the IT department of the company has full control over the device, they can lock it down to have maximum assurance that no breach of data can be carried out on this device. Furthermore, the IT department might also be responsible for repairing the device, which helps limit potential ``break-ins'' carried out during the repair process by a third party.  
	\item Responsibility for Securing the Device: The IT department of the company is fully responsible for securing the device. However, user education regarding best practices for maintaining security might still be necessary to educate individual employees. 
	\item Privacy Issues: Employees have no privacy protection. Any activity conducted using the company device could potentially be captured by the IT department -- either for operational or security reasons. 
	\item Usability and Freedom of Use for Employees: As users' access to the device would be restricted and they would not be able to install or delete applications, the freedom of use would be limited. However, usability would depend upon how the company designs its services for its employees.
\end{itemize}

It must be remembered that the COPE model is closest to the traditional model, much preferred by organisations, known as Corporate Owned Business Only (COBO) \cite{holleran2014building}. 

\subsection*{Bring Your Own Device (BYOD)}

In this paradigm, a user acquires a mobile device of her own. The user can then approach her employer and request the company application/credentials to connect her device to the aircraft system \cite{ballagas2004byod}. The mobile device management system of the company, which can be an airline, aircraft manufacturer or maintenance contractor, then enrols the given device in the system. The company can then also issue their applications to the user and assign any associated credentials. To some extent the employer can exercise limited control over the users' device. However, fundamentally the device belongs to the user and it is challenging to control a users' activity on such devices. 

Furthermore, the management of software updates to patch vulnerabilities and any additional security software on such a device might be the users' prerogative.

\begin{itemize}
	\item Device Ownership: Users purchase the device and they own it. 
	\item Application Control: The employer only has control of their own application. In contrast, the user is free to install or delete any applications they require. The employer would not be able to monitor what other applications are installed on the device. 
	\item Protecting Company Assets: Although the IT department of the employer is responsible for protecting the company's digital assets, if some sensitive data is stored on the device (as part of the company's application), it is challenging to provide assurance that such data might not be breached -- either by the user or a malicious entity who gains access to the device. 
	\item Responsibility for Securing the Device: The responsibility for updating the device in a timely manner and potentially having tools to protect against malicious intrusions falls under the purview of individual users (employees). This is a challenging prospect for the company to ensure that its employees take due care of their personal devices. User training and awareness of security concerns would be critical in this model but it has its limits. 
	\item Privacy Issues: Employers can only monitor the user activity that is carried out out using their applications or during access to their resources (i.e. company internet and website). Any other activities carried on the device are not monitored and thus provide a level of privacy to employees. 
	\item Usability and Freedom of Use for Employees: The user has full access to the device and she can use the device in the way she deems right. As before, usability deals with the ease with which she can access her company's resources -- something that is dependent upon the company's design. 
\end{itemize}

\subsection*{Choose Your Own Device (CYOD)}
Potentially a compromise between the COPE and CYOD paradigm, this is referred as the CYOD. In this model, a company defines a list of preselected and pre-configured devices that are authorised to be brought into the company's network. Employees are then give the choice of buying any one of these devices and then using them to connect with the company's network \cite{french2014current}. As the company has preselected the device and might have pre-configured it to its security requirements, it might have more trust in the device. The company can also exercise some level of security management of the device, by managing the patch updates and applications that can be installed. However, all of this is dependent upon the employee-employer relationship. A user might agree with the company managing their own application on her mobile device but not with them deciding what she can or cannot use the device for -- in this model the final owner is still the user, because she has paid for the device.

\begin{itemize}
	\item Device Ownership: Actual ownership of the device is with the employee as she has paid for it. However, the employer might exercise some privileges on such devices. 
	\item Application Control: Employees can download applications on to their devices but the employer might have some security applications installed on the device that employees cannot delete. The employee might be given a free hand to install and delete any applications they want, except the ones that the employer has installed. 
	\item Protecting Company Assets: The employer has the lead role in this but employees' assistance is necessary. Security is updated and the employer can push patches to their employee's device. Employees have to use the device in a secure way so they do not become a route to breach the company's data. 
	\item Responsibility for Securing the Device: Responsibility for this lies with both the employer and employee. 
	\item Privacy Issues: Limited privacy issues, more than BYOD but less then COPE. Company might still be able to capture the activities of the user but restrict itself to certain activities related to the security and reliability of their application. 
	\item Usability and Freedom of Use for Employees: The user has limited freedom to use the device as they desire, as long as they do not try to infringe any company policies. Usability has the same level as discussed before in COPE and BYOD. 
\end{itemize}

\subsection*{Interfacing with Aircraft Systems}
Discussion of the interface is crucial as it defines the level of control and the way in which access to the aircraft systems can be designed. There are two interfacing media; wired or wireless. In a wired medium, a mobile device connects with the aircraft system over a wired link using either a docking station or a physical port connection. Physical access is required to such connection points, which might restrict certain attackers. Furthermore, as the communication medium is wired, gaining access to communication traffic is challenging (if not impossible) because it requires, again, physical access to the communication wires. 

In contrast, if the interface is wireless then a physically restricted adversary might be able to either try to connect to the aircraft systems or at least be able to listen to what is being communicated over the wireless channel. The security requirements of the wireless and wired interfaces are starkly different for an adversary who does not need either access to the mobile device or physical access to the aircraft. 

However, for an adversary that compromises mobile devices, the notion of security in terms of wired or wireless communication channels has no implications. What this adversary is restricted to or by is the on-device security mechanisms that might prevent a malicious code/entity interfering with a sensitive process/application related to the avionics ecosystem. 

The objective of the discussion on interfacing was to get the message across that for a holistic approach to the security of mobile devices for avionics systems,  interface restrictions are an important issue, with pros and cons for both wireless and wired options. However, neither of them completely isolates nor mitigates all potential security concerns.

\section*{Evaluation Case Studies}
In the previous section, we discussed different deployment models for mobile device integration with aircraft systems. In this section, we focus on three case studies where deployment of mobile devices might be relevant. In subsequent sections we will analyse how different deployment models in the context of these case studies influence the security of mobile devices. 

\subsection*{Mobile Device as Electronic Flight Bag}
An Electronic Flight Bag (EFB) is a management device that helps flight crew to perform flight management-related tasks in an easy and convenient manner. According to the FAA Advisory Circular 120-76C, an EFB is an electronic display system intended primarily for cockpit/flight deck and/or cabin use \cite{Carrico2015}. In this paper we focus on Class 1 and Class 2 of EFB, as defined in the FAA Advisory \cite{FAA120-76C}. 

Class 1 and Class 2 EFBs are off-the-shelf mobile devices that have no FAA design, production, or installation approval, either for the whole device or its subcomponents. Devices in these two classs can connect to the aircraft system data and the only difference between them is whether the EFB is mountable or not \cite{FAA120-76C}. 

\subsection*{Mobile Devices as Maintenance Tools}
Mobile devices could allow a maintenance crew to offload the flight logs, perform diagnostics and potentially perform aircraft system configuration. Such devices might be operated either by boarding the aircraft or from the ground while the aircraft is stationary. Such devices can either connect directly with aircraft subsystems like engine management systems or may be required to be connected via an aircraft central system or hub to communicate with any of its subsystems. As noted above, these devices may have the privilege of making changes to the aircraft's operating parameters/configurations, thus making them a crucial case of mobile devices that connect with aircraft systems.

\subsection*{Mobile Devices as Operational Tools}
These mobile devices are similar to the EFBs discussed in the previous sections. However, instead of these devices being operated in the cockpit or flight deck, the cabin crew manage the inflight entertainment and ambience configuration using these devices. These mobile devices might potentially replace the setting (configuration) console in the passenger cabins that crew members currently use to control the cabin environment.

\section*{Threat Model}
In the previous section, we discussed three potential deployments of mobile devices as part of a digital avionics network. These three deployment models have their own security and reliability concerns, which maybe unique to each of their deployment scenarios and applications/data they deal with. However, in this section we discuss two generic types of adversary in the context of mobile devices as defined in this paper. It must be remembered that this categorisation is made based on the level of access each adversary has to the mobile device, not on the basis of individual adversaries' capabilities to compromise a given device.  

\subsection*{On-Device Adversary}
On-device adversaries are malicious users that can potentially compromise a mobile device. By doing so, depending upon the depth of compromise, they can control the execution of any sensitive applications that connect with the aircraft system and communication of relevant information. The closer an adversary can compromise the layer to the hardware of the mobile device, the more powerful he or she is -- with abilities to interrupt or modify the execution of and/or data communication from a sensitive application running on the device. 

\begin{figure}[htbp]
	\centering
		\includegraphics[width=1.00\columnwidth]{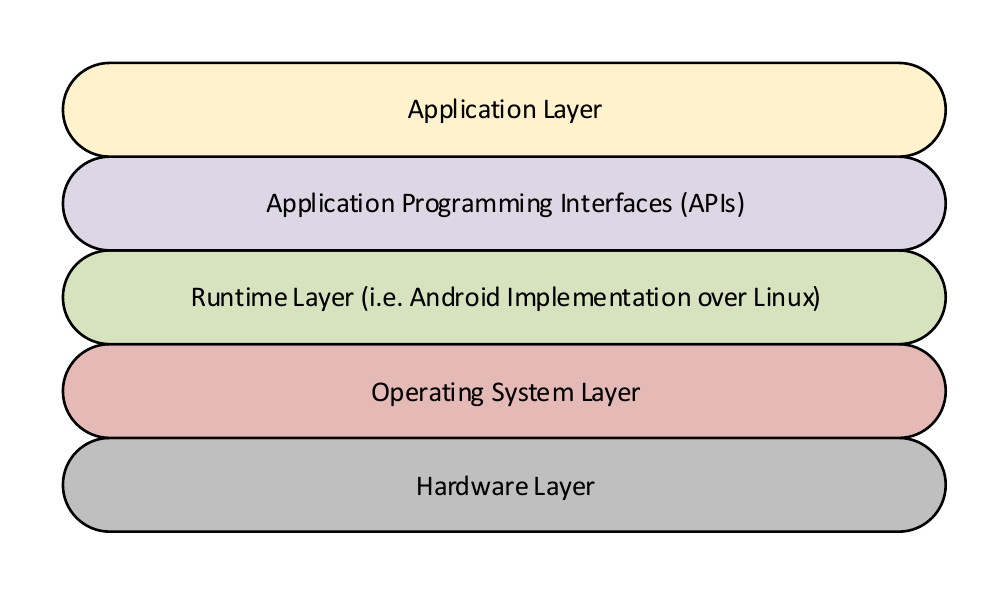}
	\caption{On-Device Adversary Target Layers}
	\label{fig:On-boardAdversary}
\end{figure}

For an on-device adversary, the aim is to compromise any of the layers (semantically) represented in Figure \ref{fig:On-boardAdversary}. The lower the layers an adversary manages to compromise, the stronger the potential to influence an application and its data.  The adversary may have compromised the device either at the manufacturing stage, by getting physical access even for a short while, or remotely, by installing a malicious application on the device. Such an adversary might be physically distant from the compromised device but potentially still be able to control it.

\subsection*{Off-Device Adversary}
Off-device adversaries are malicious users that have no access to the mobile devices -- either physical or remote. These adversaries try to intercept the communications between a mobile device and the aircraft system. In the interface section we mentioned that it is comparatively easy to intercept a wireless channel compared to a wired channel, but for simplicity we assume that an off-device adversary has equal access to all communication media between a mobile device and the aircraft system.

Unlike the on-device adversary, the off-device adversary has to be near the geographic location of the mobile device to effectively intercept its communication. This adversary does not need to compromise either the mobile device or the aircraft systems -- they only need to have the ability to intercept the communications between the mobile device and the aircraft system.

\section*{Evaluation of Mobile Devices as Digital Avionics Component}
In this section, we briefly evaluate the risk each of the mobile device case studies might face from the two generic adversaries discussed in the previous section. In this section we are not evaluating any particular product that actually deploys mobile devices in the respective case studies but are considering a generic high-level analysis of what security risks these deployments might face. 

\subsection*{Mobile Device as Electronic Flight Bag}
Mobile devices such as EFBs may have strong access control and firewalls on the aircraft system side, which may prevent a non-authorised application from connecting and communicating potentially incorrect data. However, an on-device adversary can potentially control the execution of the sensitive application that is authorised to communicate with the aircraft system. Any cryptographic mechanism deployed for either authentication or encryption/signature in the data communicated from the device to the aircraft would potentially be vulnerable to an on-device adversary. Therefore, any security mechanism designed as part of the sensitive mobile application would not be effective if the adversary has access to the runtime environment, or worse, access to the hardware. It should be noted that any mechanism built on the aircraft side of this would have little to no effect in preventing an on-board adversary from interfering with sensitive applications activity.

Theoretically, if off-device adversaries gain access to the communications between the mobile device and aircraft systems, they might be able to eavesdrop. An active off-device adversary might even be able to inject some data into this communication; however, data integrity is not assured in this communication.

\subsection*{Mobile Devices as Maintenance Tools}
Similar to EFBs, mobile devices as maintenance tools might have few defences against on-board adversaries if any or all countermeasures implemented by the application developer are based on the applications. However, it should be noted that other mechanisms also do not fare well against an advanced adversary that has the ability to compromise the mobile device. If the aircraft configuration is generated on a mobile device and then uses secure cryptographic mechanisms communicated to the aircraft systems, such an update would not be secure against an on-device adversary. However, if the mobile device is just a communication conduit between the back-office system and the aircraft system and all updates are secured end-to-end, an on-device adversary would not be able to modify the end-to-end communication channel. 

In contrast, for the off-device adversary, if the communication is unencrypted or there is a vulnerability in the communication scheme then he or she might be able to potentially exploit it. However, as an off-device adversary does not have the capability to break the standard cryptographic algorithms, which means he or she cannot break into encrypted communication. Thus creating a strong limitation on what such adversaries can achieve. 

\subsection*{Mobile Device as Operational Tool}
Unlike the previous two cases, it can be argued that mobile devices as operational tools pose the least amount of safety risk to aircraft systems. As discussed in the previous sections, an operational tool might deal with the in-flight entertainment and/or cabin ambience. The maximum an attacker could achieve would be modifying some parameters in such systems -- if such a provision were allowed in the system. The on-device adversary, by now recognised to be the most advanced adversary, can manage to interrupt execution and potentially can process any command as desired, whereas the off-board adversary can mount replay attacks, man-in-the-middle attacks and/or denial of service attacks. Of these, replay attacks may pose the most significant danger as they could trigger certain cabin conditions, which might endanger the cabin passengers. 

\section*{Guidelines for a Secure and Trusted Integration}
In the previous sections we discussed two categories of adversaries, their capabilities and their risk potential. The question that arises is ``How real is the threat of the on-board adversary?'' A valid question and the potential of it being realised is not the realm of fantasy. There are examples in which devices were shipped with built-in Trojans and/or an advanced adversary compromised devices after being issued to users \cite{la2013survey}.  As aircraft safety is paramount and its reliability is stringently required, we consider that designing a system that is secure against such adversaries is the only reasonable option. In this section, we list some of the security considerations that should be taken into account when deploying mobile devices to interface and communicate with aircraft systems. 

\subsection*{Least Privilege Architecture}
Each user and mobile device that connects with the aircraft system should be authenticated. The access privilege issued to the user and mobile device should be atomic. An atomic access is defined as individual users and devices having their own sets of separate privileges -- unique to each user and device. The privileges given to a user will only be those that the device is also permitted to have. Furthermore, each access privilege should have a clear and pre-defined time scale; after this time the user and device should re-authenticate themselves to the aircraft systems. 

Certain aircraft systems should only allow access when certain environmental conditions are met. For example, maintenance crew might only be allowed to access maintenance logs once the aircraft is grounded and stationary. Furthermore, a reconfiguration of an aircraft system is only allowed if the aircraft is grounded and in maintenance phase. 

\subsection*{Trusted and Isolated Enrolment}
Mobile devices and users have to be enrolled in the relevant organisation's privilege system to gain access to aircraft systems. These organisations can be airline operators, aircraft maintainers and manufacturers. In all cases, the mobile device and user enrolment should be closely monitored. There should be a robust system to vet users and devices (if possible) at this stage.

\subsection*{Strong Function Classification}
Each application issued to a user and a device should have a restricted code base. Only the application code that is necessary to perform the relevant tasks should be visible. Hiding functionality based on access privilege is not a preferred solution in this case. 

\subsection*{Strong Binding with Aircraft and Operational-Environment}
Each device and user account should be associated with an individual aircraft. This is to avoid potential issues in which user or device credentials are used to access an aircraft system when the person is nowhere near the aircraft or not working at the time of access. The access credentials and privilege to access an aircraft should be as unique as possible and as restricted in time and geographical location as possible. Furthermore, mobile devices (if possible) should have a strong binding to the aircraft systems.

\subsection*{Hardened Firewall Mechanism}
A strong firewall mechanism with an in-depth packet inspection scheme should be considered for mobile connectivity with aircraft systems. Furthermore, firewalls should also be able to detect any covert channels and enforce strong information flow policies. 

\subsection*{Strong Access Control Mechanism}
Access control should be implemented on strong authentication schemes. Both user and device should be authenticated separately. User authentication is based on stronger mechanisms like biometrics (widely available on mobile devices now) and device authentication should be based on two-way challenge-response protocols.

\subsection*{Strong Secure and Trusted Channels}
Any communication between a mobile device and the aircraft system, whether via a wired or wireless interface, should be protected using cryptographic mechanisms. For this purpose, a secure and trusted channel protocol should be deployed. In a secure and trusted channel protocol, not only do the communicating entities authenticate to each other but also their internal states are validated as trustworthy. For validation of the software (and potential hardware) state of the aircraft application on the mobile device, a trusted platform architecture could be deployed -- this is discussed in subsequent sections. This will not only ensure that the communication channel is protected and devices are authenticated but also that the status of applications on the devices are also secure (and free of any malicious alterations). For an in-depth analysis and security recommendations on how to design a secure channel for digital avionics systems, please refer to \cite{akram2015challenges}.

\subsection*{Data Integrity, Traceability, and Validation}
Any data loaded or off-loaded from an aircraft should provide strong integrity, traceability and validation properties. A strong integrity mechanism provides an assurance that data is not being modified by any non-authorised entity during its storage and transit. For integrity mechanisms, cryptographic primitives like hash functions and digital signatures can be deployed. Data traceability provides a mechanism in which data can be traced from its creation to destruction. Such a mechanism is necessary for data quality and forensics purposes. Data validation is a mechanism in which certain elements of data are created in a specific way, or certain errors are left in the data in a manner that allows a trusted entity to verify the origin of data. This mechanism can ascertain whether data presented to the aircraft or data from an aircraft to the airline/maintenance back office can be validated as being generated by the entity to which it is attributed. 

\subsection*{Trusted Platform}
There are many proposals that promote trusted platform architectures. In this section we will discuss three of these:

\subsubsection*{Trusted Platform Module}
The definition of trust, taken from Merriam Webster's online dictionary\footnote{Website: \url{http://www.merriam-webster.com/dictionary/trust}} is that trust is a ``belief that someone or something is reliable, good, honest, effective, etc."

The TPM specifications are maintained and developed by an international standards group called the Trusted Computing Group (TCG)\footnote{Trusted Computing Group (TCG) is the culmination of industrial efforts that included the Trusted Computing Platform Association (TCPA), Microsoft's Palladium, later called Next Generation Computing Base (NGSCB), and Intel's LaGrande. All of them proposed how to ascertain trust in a device's state in a distributed environment. These efforts were combined in the TCG specification that resulted in the proposal of TPM.} Today, TCG not only publishes the TPM specifications but also the Mobile Trusted Module (MTM), Trusted Multi-tenant Infrastructure, and Trusted Network Connect (TNC). Facing emerging technologies, service architectures, and computing platforms, TCG is adapting to the challenges presented by them. 

The TPM chip, whose specification is defined by the Trusted Computing Group \cite{TCG}, is known as the hardware root-of-trust into the trusted computing ecosystem. Currently it is deployed to laptops, PCs, and mobiles and is produced by manufacturers including Infineon, Atmel and Broadcom. At present, the TPM is available as a tamper-resistant security chip that is physically bounded to the computer's motherboard and controlled by software running on the system using well-defined commands. The TPM MOBILE with Trusted Execution Environment has recently emerged; its origin lies in the TPM v1.2 a with some enhancements for mobile devices \cite{TPMSpec2011} . 
The TPM provides:

\begin{enumerate}
\item Roots of Trust include hardware/software components that are intrinsically trusted to establish a chain of trust that ensures only trusted software and hardware can be used (see the Mobile Trusted Module (MTM) section).
\item The Platform Configuration Register (PCR) in the most modern TPMs includes 24 registers. It is used to store the state of system measurements. These measurements are represented normally by a cryptographic hash computed from the hash values (SHA-1) of components (applications) running on the platform. PCRs cannot be written directly; a process called extending the PCR can only store data.
\item The RSA keys: There are two types of RSA keys that TPM generates and which are considered as root keys (they never leave the TPM): 
\begin{enumerate}
\item Endorsement Key (EK): This key is used in its role as a Root of Trust for Reporting. During the installation of an owner in the TPM, the manufacturer generates this key with a public/private key pair built into the hardware. The public component of the EK is certified by an appropriate CA, which assigns the EK to a particular TPM. Thus, each individual TPM has a unique platform EK. 
For the private component of the EK, the TPM can sign assertions about the trusted computer's state. A remote computer can verify that those assertions have been signed by a trusted TPM. 
\item Storage Root Key (SRK): This key is used to protect other keys and data via encryption. 
\item Attestation Identity Keys (AIKs): The AIK is used to identify the platform in transactions such as platform authentication and platform attestation. Because of the uniqueness of the EK, the AIK is used in remote attestation by a particular application. The private key is non-migratable and protected by the TPM and the public key is encrypted by a storage root key (or other key) outside the TPM with the possibility to be loaded into the TPM. The security of the public key is bootstrapped from the TPM's EK.
The AIK is generally used for several roles: signing/reporting user data; storage (encrypting data and other keys); and binding (decrypting data, used also for remote parties).
\end{enumerate}
\end{enumerate}

\subsubsection*{Trusted Execution Environment}
A Trusted Execution Environment (TEE) provides the necessary assurance that during the execution of an application, no on-board application can interfere with its execution. Two of the main proposals for the TEE are ARM TrustZone and GlobalPlatform TEE -- in recent years these two proposals have been converging but in this section we have briefly discussed them separately. The ARM TrustZone also provides the architecture for a trusted platform specifically for mobile devices. The underlying concept is the provision of two virtual processors with hardware-level segregation and access control \cite{10.1109/MDT.2007.196,ARMTrustZone2009}. This enables the ARM TrustZone to define two execution environments described as Secure world and Normal world. The Secure world executes the security- and privacy-sensitive components of applications and normal execution takes place in the Normal world. The ARM processor manages the switch between the two worlds. The ARM TrustZone is implemented as a security extension to the ARM processors (e.g.\ ARM1176JZ(F)-S, Cortes-A8, and Cortex-A9 MPCore) \cite{ARMTrustZone2009}, which a developer can opt to use if required.

\begin{figure}[htbp]
\centering
\includegraphics[width=0.99\columnwidth]{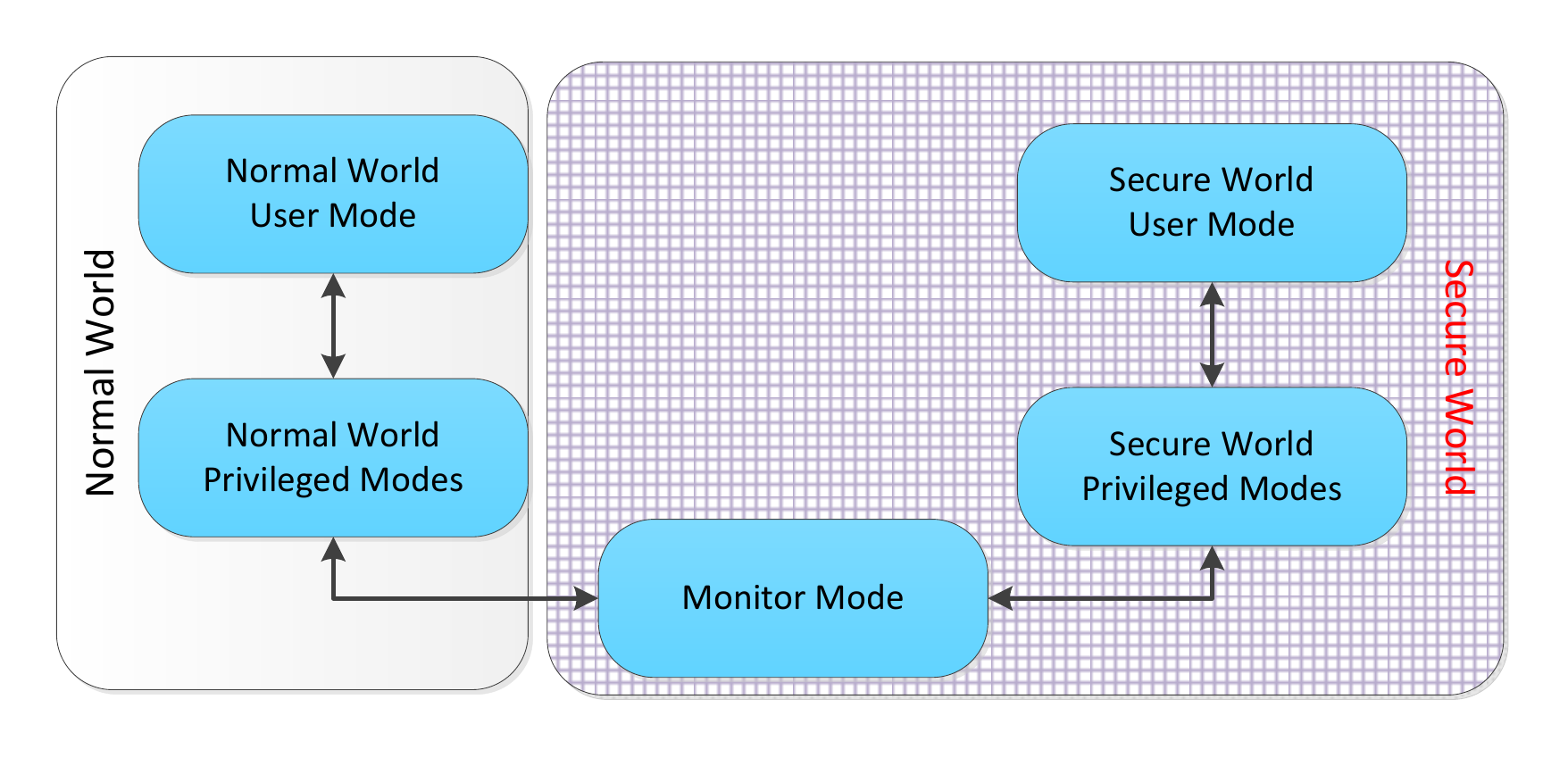}
\caption{Generic architectural view of ARM TrustZone}
\label{fig:MTMOperations}
\end{figure}

The TEE is GlobalPlatform's initiative \cite{GPDSTIP2007, GP:DASM2008a, TEE2011} for mobile phones, set-top boxes, utility meters, and payphones. GlobalPlatform defines a specification for interoperable secure hardware, which is based on GlobalPlatform's experience in the smart card industry. It does not define any particular hardware, which can be based on either a typical secure element or any of the previously discussed tamper-resistant devices. The rationale for discussing the TEE as one of the candidate devices is to provide a complete picture. The underlying ownership of the TEE device still predominantly resides with the issuing authority, which is similar to GlobalPlatform's specification for the smart card industry \cite{2006}.

\subsubsection*{Encrypted Execution}
An application is executed in a manner such that all of the application instructions on persistent and non-persistent storage are in encrypted format \cite{RobPaperCPSS2016}. The application is executed on processes that decrypt the application n execution cycles before the instruction is going to be executed -- n has to be as small as possible to provide efficient performance. Such a solution considers that an adversary has control over the software and hardware of a device except for the internal circuitry of a processor, which is considered to be trusted. This proposal relies on security and trust in the utmost basic element of computing -- a processor. If we consider that an adversary has a compromised processor then it is difficult to hide application execution from it. 

\subsection*{Secure and Trusted Supply Chain for Mobile Applications}
Provisioning of the application to users mobile device should be closely managed and monitored. Applications that could have the capability to be connected to an aircraft should only be available to the mobile device via a restricted supply chain. This means that such an application should not be accessible on general-purpose application distribution channels. 

\subsection*{Secure Decommissioning}
At the end of the lifecycle of the mobile device or the presence of an application on such a device, the application and device should be properly decommissioned. Recycling should be carried out in such a manner that all security-related parameters are completely removed and the device itself is put on a black list maintained by the airline or maintenance organisation.

\section*{Conclusion}
In this paper, we looked into the provisioning of connecting mobile devices in different operational capacities with aircraft systems. It can be argued that mobile devices have the potential to provide benefits; however, in this paper we investigated the security implications of such an operational situation. 

We have described three different mobile device integration models -- COPE, BYOD, and CYOD. We have also provided a comparison between these three models based on security control and responsibilities. With these integration models an important aspect is how a device can interface with an aircraft, which is either via a wired or wireless interface. In this paper, we have considered three deployed case scenarios and potential threat models are presented. We have divided potential adversaries into two categories -- on-board and off-board adversaries. It is apparent that an on-board adversary has more capacity to cause harm than an off-board adversary, but this in no way means that when designing such an integration, we can ignore the off-board adversary.

We then briefly evaluated the case scenarios based on the adversary's capabilities. Based on this analysis we have presented a minimum set of guidelines to be followed when mobile devices are considered for integration with an aircraft system.

\section*{Acknowledgements}

The authors acknowledge the support of the UK's innovation agency, InnovateUK, and the contributions of the Secure High-Availability Avionics Wireless Networks (SHAWN) project partners.  

\section*{Disclaimer}
The views and opinions expressed in this article are those of the authors and do not necessarily reflect the position of the SHAWN project or any of organisations associated with this project. 

\printbibliography

\vspace{0.5cm}
\centering
\emph{\large 2016 Integrated Communications Navigation \\and Surveillance (ICNS) Conference\\
April 19-21, 2016}

\end{document}